\documentclass[3p,times,procedia]{elsarticle}

\usepackage{ecrc}

\volume{00}

\firstpage{1}

\journalname{Physics Procedia}

\runauth{J. Yamaoka et al.}

\jid{phpro}

\jnltitlelogo{Physics Procedia}

\CopyrightLine{2011}{Published by Elsevier BV. Selection and/or peer-review under responsibility of the organizing committee for TIPP 2011.}

\usepackage{amssymb}

\usepackage{graphicx}
\DeclareGraphicsExtensions{.pdf}

\usepackage{subfig}
\usepackage{float}

\usepackage[figuresright]{rotating}

\begin{document}

\begin{frontmatter}



\dochead{TIPP 2011 -  Technology and Instrumentation in Particle Physics 2011}


\title{Application of Time Projection Chambers with GEMs and Pixels to WIMP Searches and Fast Neutron Detection}


\author[uh]{J. Yamaoka\footnote{Email: yamaoka@phys.hawaii.edu}}
\author[uh]{H. Feng}
\author[lbnl]{M. Garcia-Sciveres}
\author[uh]{I. Jaegle}
\author[lbnl]{J. Kadyk}
\author[lbnl]{Y. Nguyen}
\author[uh]{M. Rosen}
\author[uh]{S. Ross}
\author[uh]{T. Thorpe}
\author[uh]{S. Vahsen}

\address[uh]{University of Hawaii at Manoa, 2505 Correa Rd., Honolulu, HI 96822}
\address[lbnl]{Lawrence Berkeley National Laboratory, 1 Cyclotron Rd., Berkeley, CA 94720}

\begin{abstract}
We present work on the detection of neutral particles via nuclear recoils in gas-filled Time Projection Chambers (TPCs).  We employ Gas Electron Multipliers (GEMs) to amplify the signal and silicon pixel electronics to detect the avalanche charge.  These technologies allow ionization in the target gas to be detected with low noise, improved position and time resolution, and high efficiency. We review experimental results obtained in previous years, and report on ongoing simulation studies and construction of the first prototype at the University of Hawaii.  We also present prospects of using such detectors to perform direction-sensitive searches for WIMP dark matter and fast neutron from fissionable material.
\end{abstract}


\end{frontmatter}


\section{Introduction}
\label{Intro}

\subsection{Time Projection Chambers}
\label{TPC}

Time Projection Chambers (TPCs)~\cite{Marx:1978zz}, originally invented in 1976, have been used for many years for particle tracking in high energy physics experiments.  Typically the readout of the ionization in the chamber is achieved with a large array of wires on one end which must be both electrically and mechanically stable.  As proposed in~\cite{Kim:2008zzi}, we are attempting to update the readout of the TPC with a modern scalable solution.  We replace the wire array with the ATLAS FE-I3 pixel chip~\cite{Peric:2006km}.  Here we are able to leverage the development done by ATLAS to implement a fully realized system (including triggering and readout) with a well established production chain.  Large numbers of these chips can be used in arrays to read out large volume TPCs.  This is discussed in Section~\ref{Fut_Plans}.  We also employ Gas Electron Multipliers (GEMs)~\cite{Sauli:1997qp}, which avalanche (amplify) the ionized electrons, to achieve high efficiency, even for single electrons. 

\subsection{Physics Motivation}
\label{Phys-Mot}

\begin{figure}[htb]
 	\centering
  	\subfloat[1 MeV H recoils from neutrons in $C_{4}H_{10}$ gas at 1 ATM]{\label{fig:scrim_neu}\includegraphics[width=0.4\textwidth]{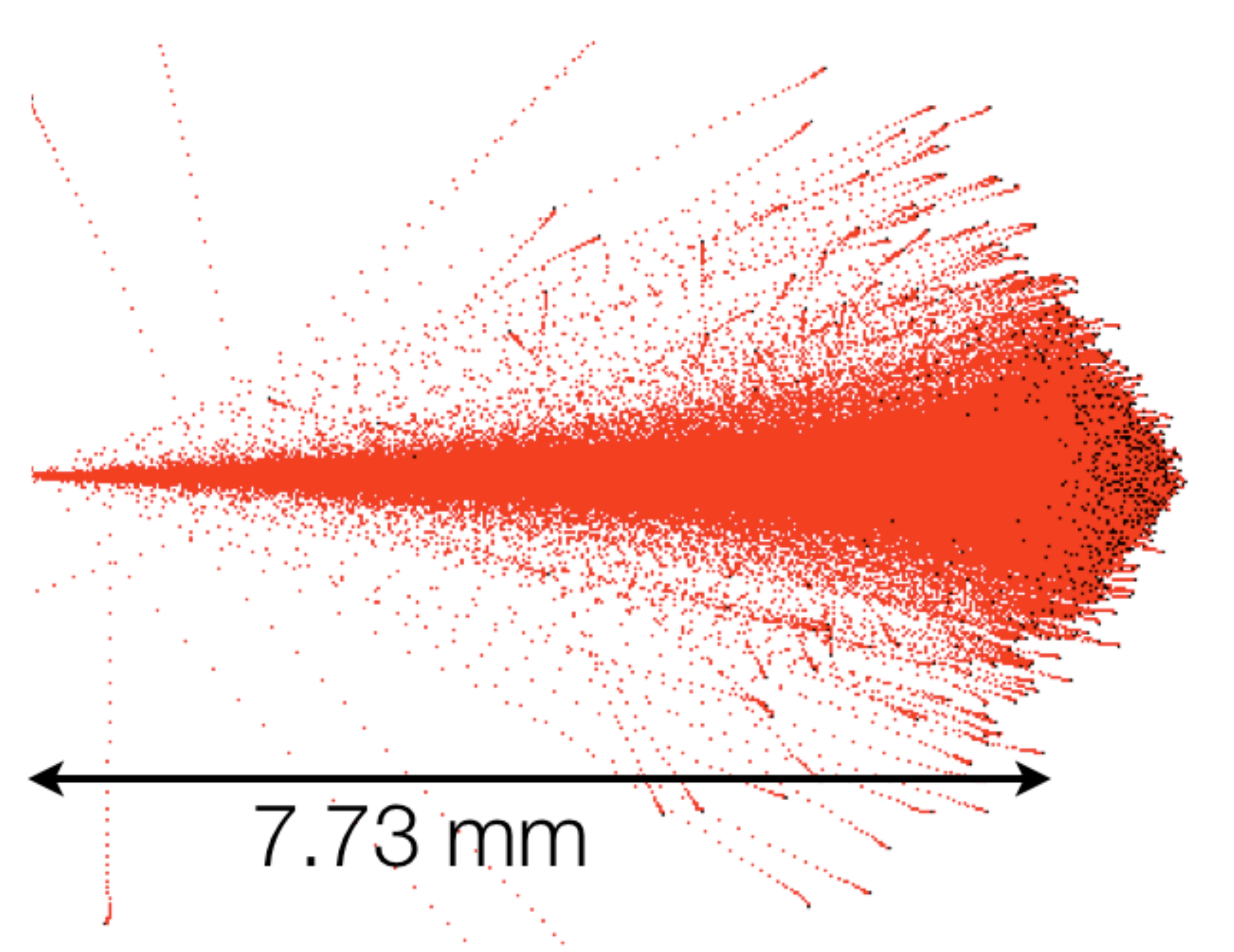}}                
	\hspace{0.1\textwidth}
  	\subfloat[100 keV F recoils from WIMPs in $CF_{4}$ at 75 Torr]{\label{fig:scrim_wimp}\includegraphics[width=0.4\textwidth]{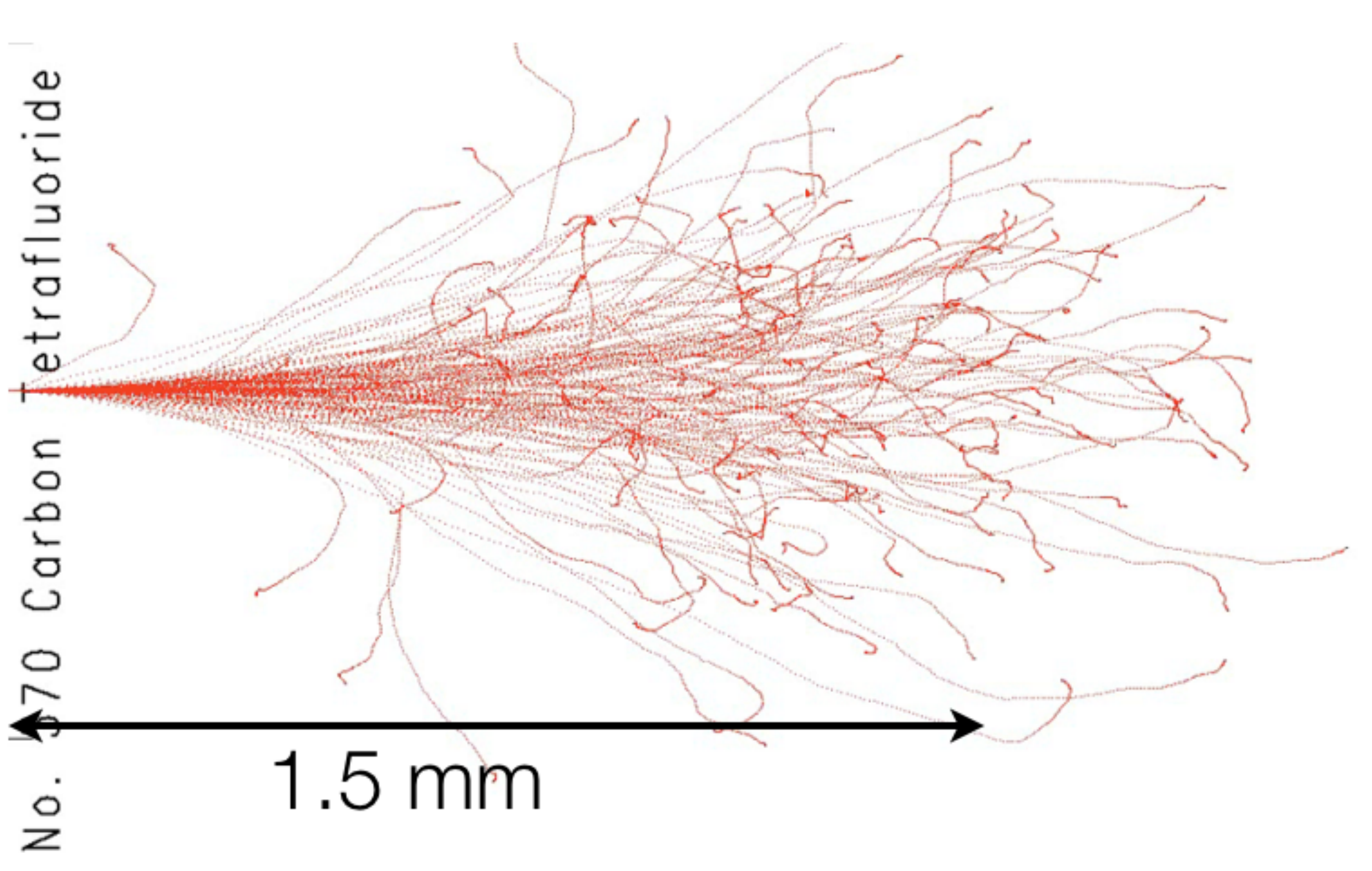}}
  	\caption{{\tt SRIM} simulation of nuclear recoils in gas from neutron (a) and WIMPs (b).}
  	\label{fig:scrim}
\end{figure}

Because of the excellent spacial ($\sim$$14~\mu m$) and timing ($25~ns$) resolution of the ATLAS pixel chip, we should be able to reconstruct relatively short tracks in three dimensions. Therefore, this type of detector should be effective for detecting neutral particles via the tracks produced in elastic scattering off the nuclei of the gas.  These types of interactions can be used to detect particles such as neutrons or perhaps weakly interacting dark matter particles (WIMPs).  Figure~\ref{fig:scrim}, shows nuclear recoils simulated with {\tt SRIM}~\cite{srimtext} for neutrons and WIMPs.  The additional ability of the chip to measure the total charge should also enable us to determine the incoming direction of the particle.  

One striking signature of dark matter would be the twelve hour directional oscillation of the WIMP recoils as the Earth rotates~\cite{Ahlen:2009ev}.  There is also interest to use such a detector to replace $^{3}He$ based neutron detectors used to detect nuclear material.  Though not discussed here, similar TPCs are being proposed to study beam backgrounds at the Super KEKB accelerator in Japan.

\section{Detector Prototypes}
\label{Det}

\subsection{First Generation}
\begin{figure}[htb]
	\centering
        \includegraphics[width=0.8\textwidth]{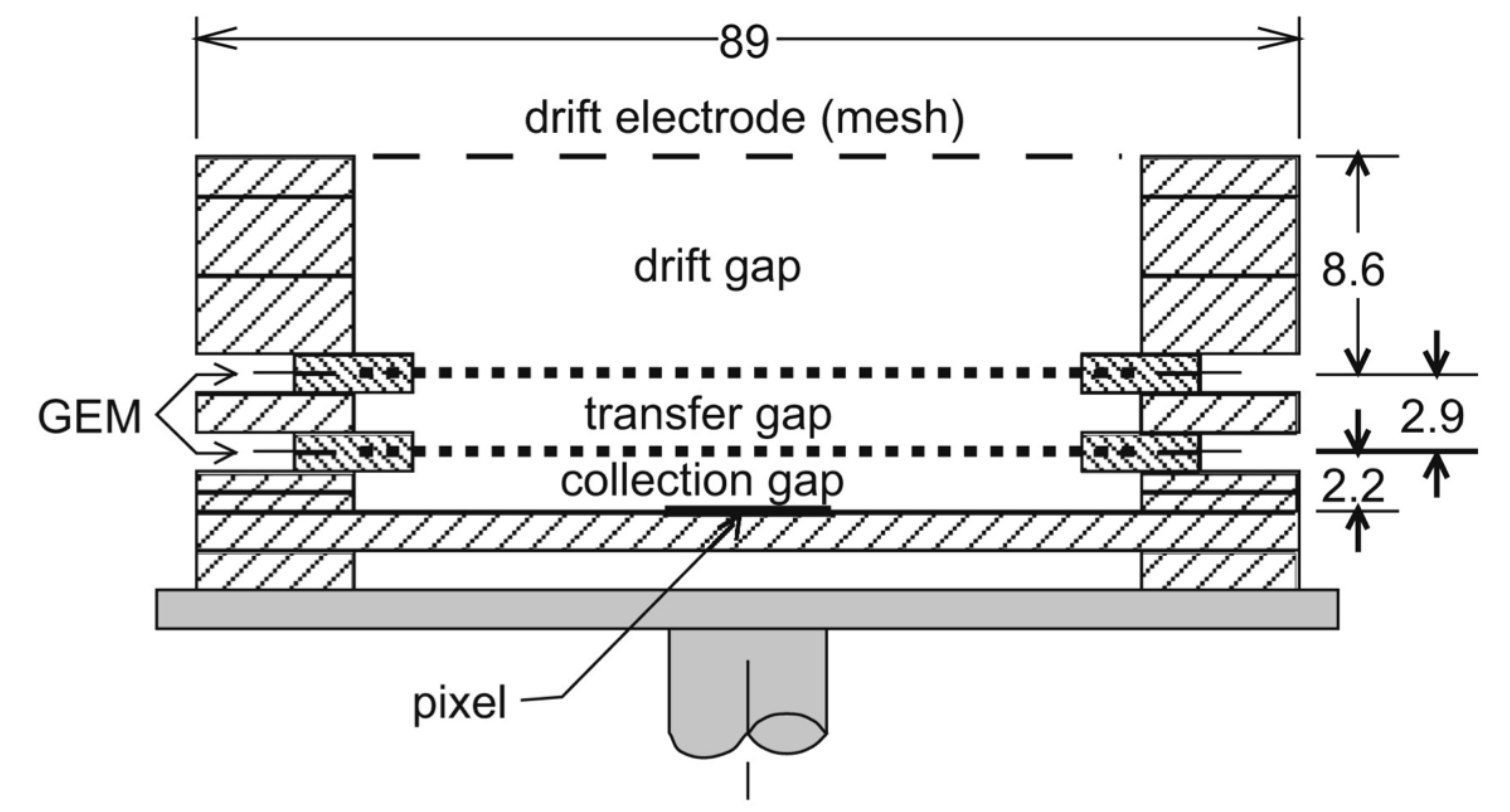}
        \caption{Cross section of GEM and pixel chip fixture used in the First Generation prototype~\cite{Kim:2008zzi}.  Dimensions in mm.}
        \label{fig:schim1}
\end{figure}

The construction of the first generation prototype was led by J. Kadyk at Lawrence Berkeley National Laboratory (LBNL)~\cite{Kim:2008zzi}.  This prototype consists of a small drift gap with a drift electrode (consisting of a wire mesh) above and two layers of GEMs and the pixel chip below, Figure~\ref{fig:schim1}.  This apparatus is contained within a vacuum vessel filled with a target gas.  Several gases were studied ($Ar/CO_2$, $Ar/C_4H_{10}$, and $Ar/CH_4/CO_2$), and more work will be done to optimize the gas for WIMPs or neutrons, but the results shown are with $Ar/CO_2$ in a ratio of 70/30 at 1 ATM.

\subsubsection{ATLAS FE-I3 Pixel Chip}
\begin{figure}[htb]
 	\centering
  	\subfloat[FE-I3 pixel chip on ATLAS test board.]{\label{fig:pixchip}\includegraphics[width=0.4\textwidth]{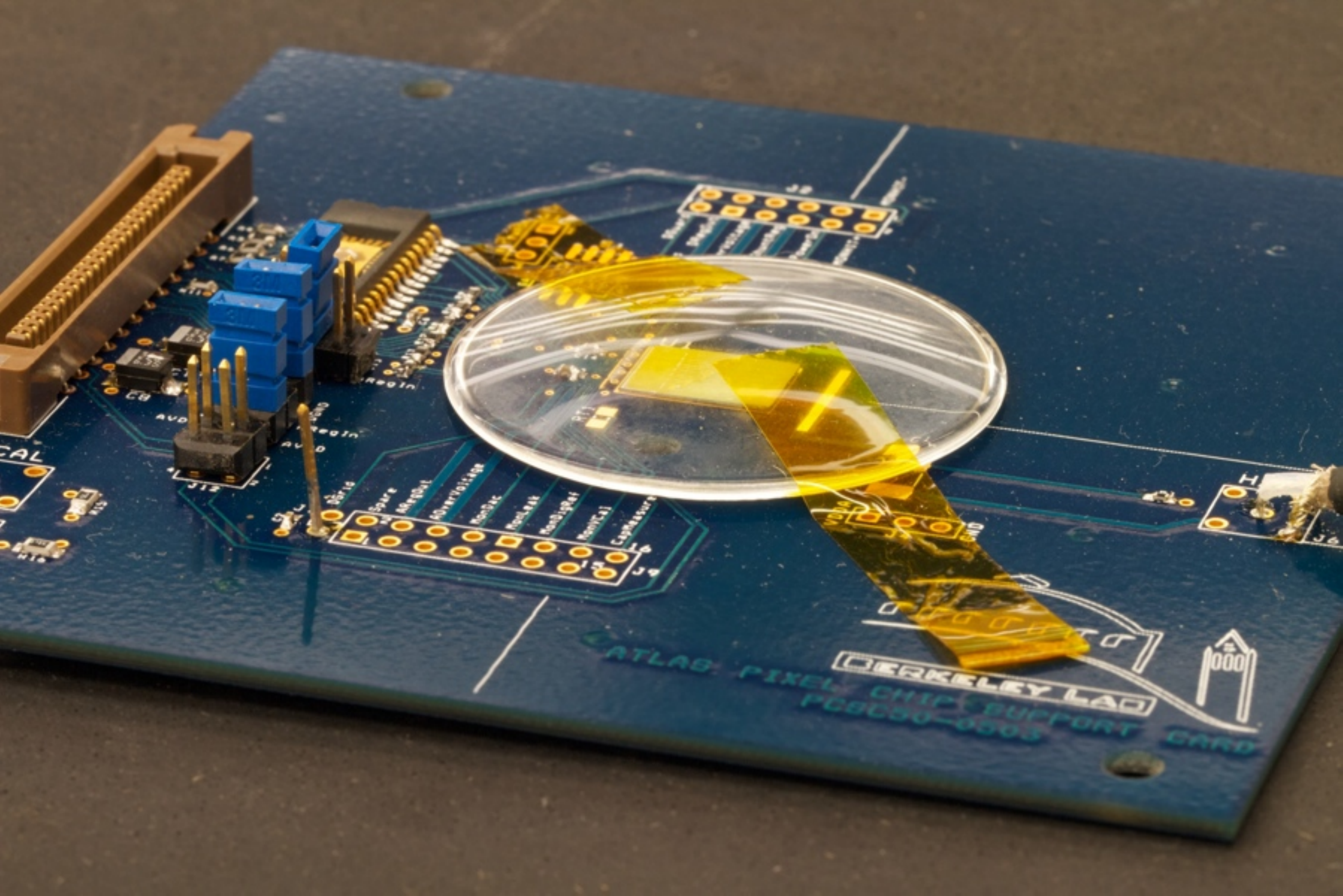}}                
	\hspace{0.1\textwidth}
  	\subfloat[Modified pixel board suitable for use in TPCs~\cite{Kim:2008zzi}]{\label{fig:pix_tpc}\includegraphics[width=0.4\textwidth]{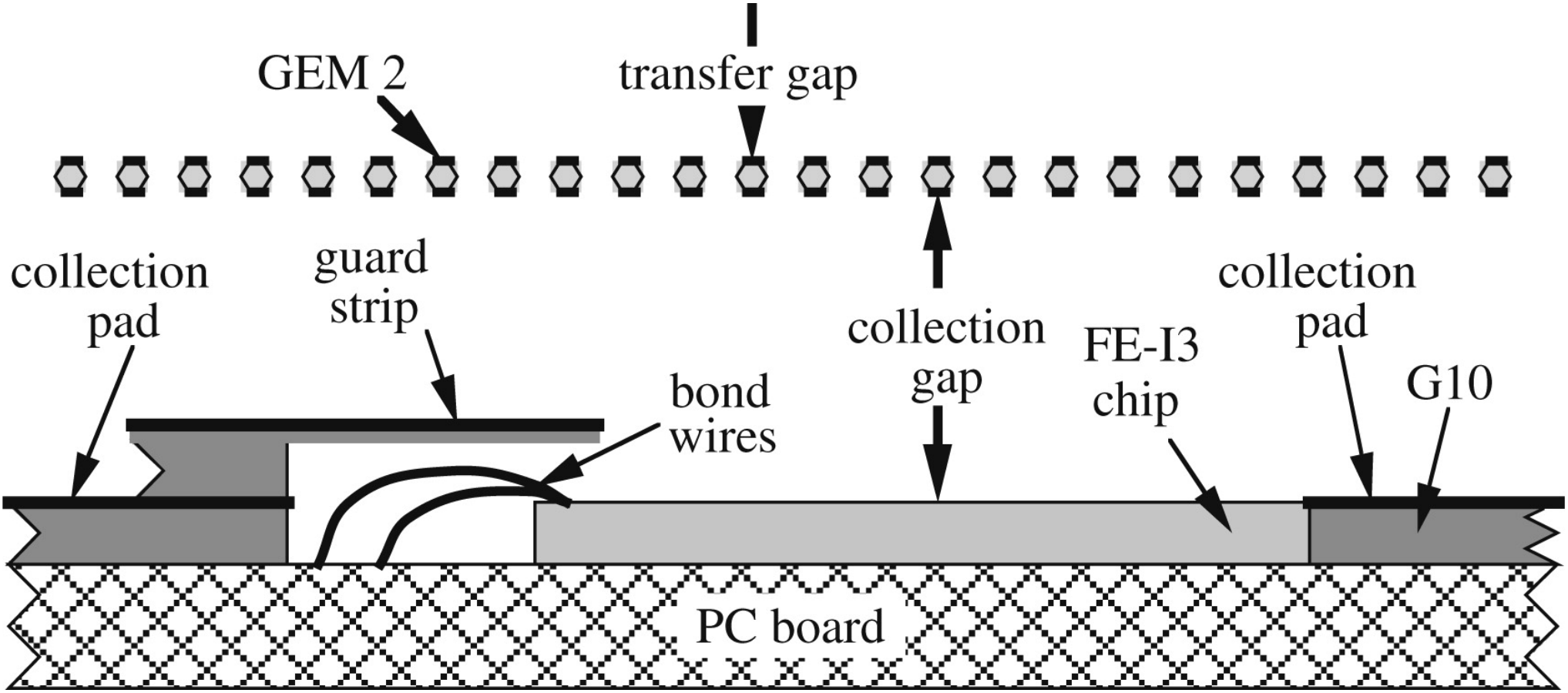}}
  	\caption{ATLAS test board and schematic of modified board for use in TPCs.}
  	\label{fig:pix}
\end{figure}

The ATLAS FE-I3 pixel chip (currently used in the ATLAS detector) was developed at LBNL over a 7 year period.  It has an active area of 7.2 mm x 10.8 mm, with 2880 pixels of dimensions 50 $\mu m$ x 400 $\mu m$ arranged into 18 columns and 160 rows.  The chip has very low noise, less than 120 electrons per pixel.  The chip is self-triggering (with an adjustable trigger threshold) and has a readout integration time of 400 ns.  In $Ar/CO_2$ with a drift field 1 kV/cm, this equates to $\sim$10 mm. 

Because of the requirements of TPCs, some slight modifications of the standard ATLAS pixel chip test board, Figure~\ref{fig:pixchip}, are needed.  The pixel sites on the FE-I3 chip were plated with either gold or aluminum to help maintain a uniform electric field and to help prevent surface charge from from building on the semiconductor.  The board was redesigned to move the surface mounted readout electronics to the back of the board, and a copper collection pad and guard strip were also added, see Figure~\ref{fig:pix_tpc}. 

\subsubsection{Gas Electron Multipliers (GEM)}
\label{ss:gems}

\begin{figure}[H]
 	\centering
  	\subfloat[GEM simulation with {\tt GARFIELD}~\cite{Veenhof:1998tt}, with 500 V GEM bias.]{\label{fig:gemsim}\includegraphics[width=0.45\textwidth]{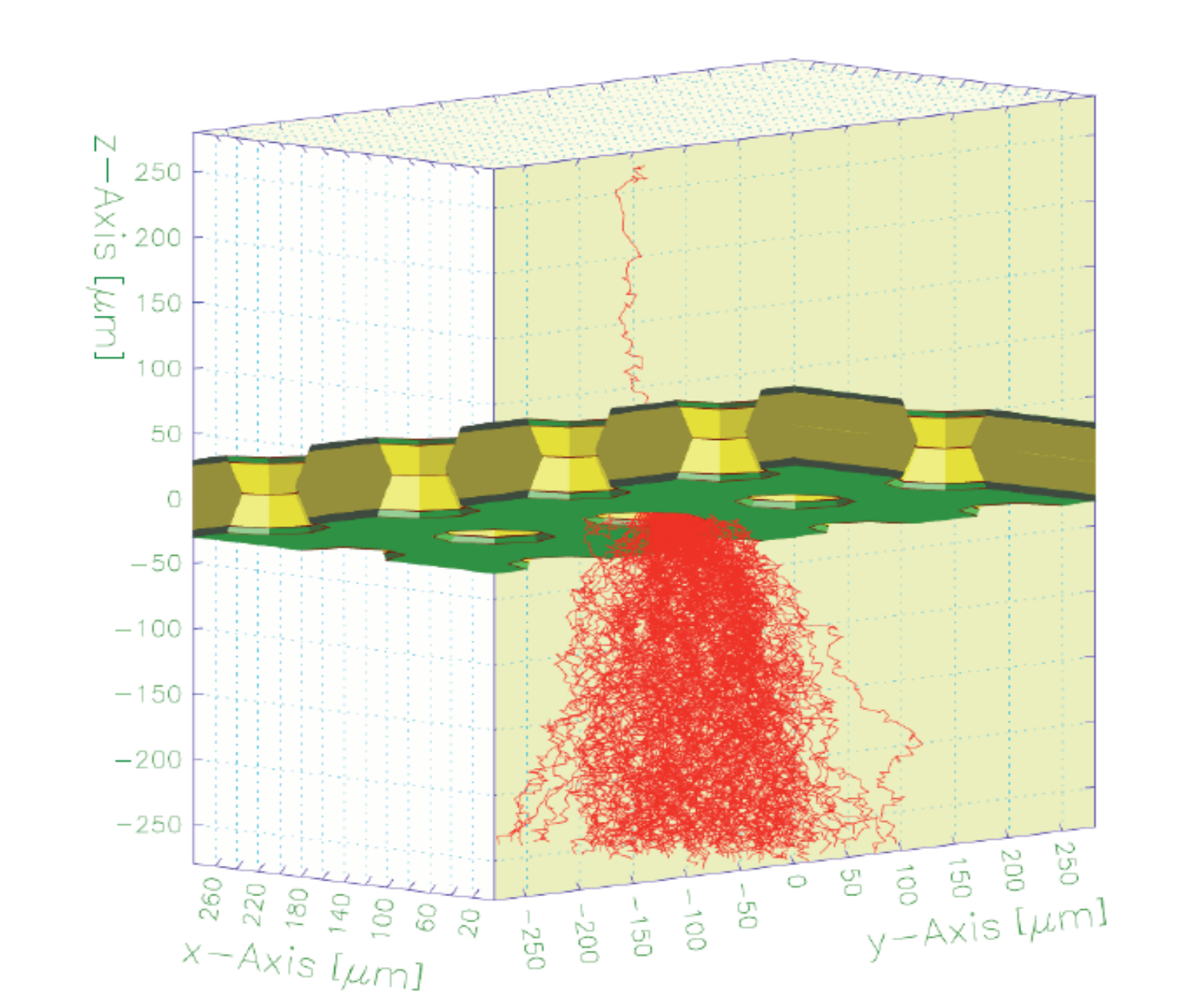}}                
	\hspace{0.1\textwidth}
  	\subfloat[Energy spectrum~\cite{Kim:2008zzi} for $^{55}Fe$.]{\label{fig:espec}\includegraphics[trim = 1mm 10mm 1mm 0mm, clip,width=0.35\textwidth]{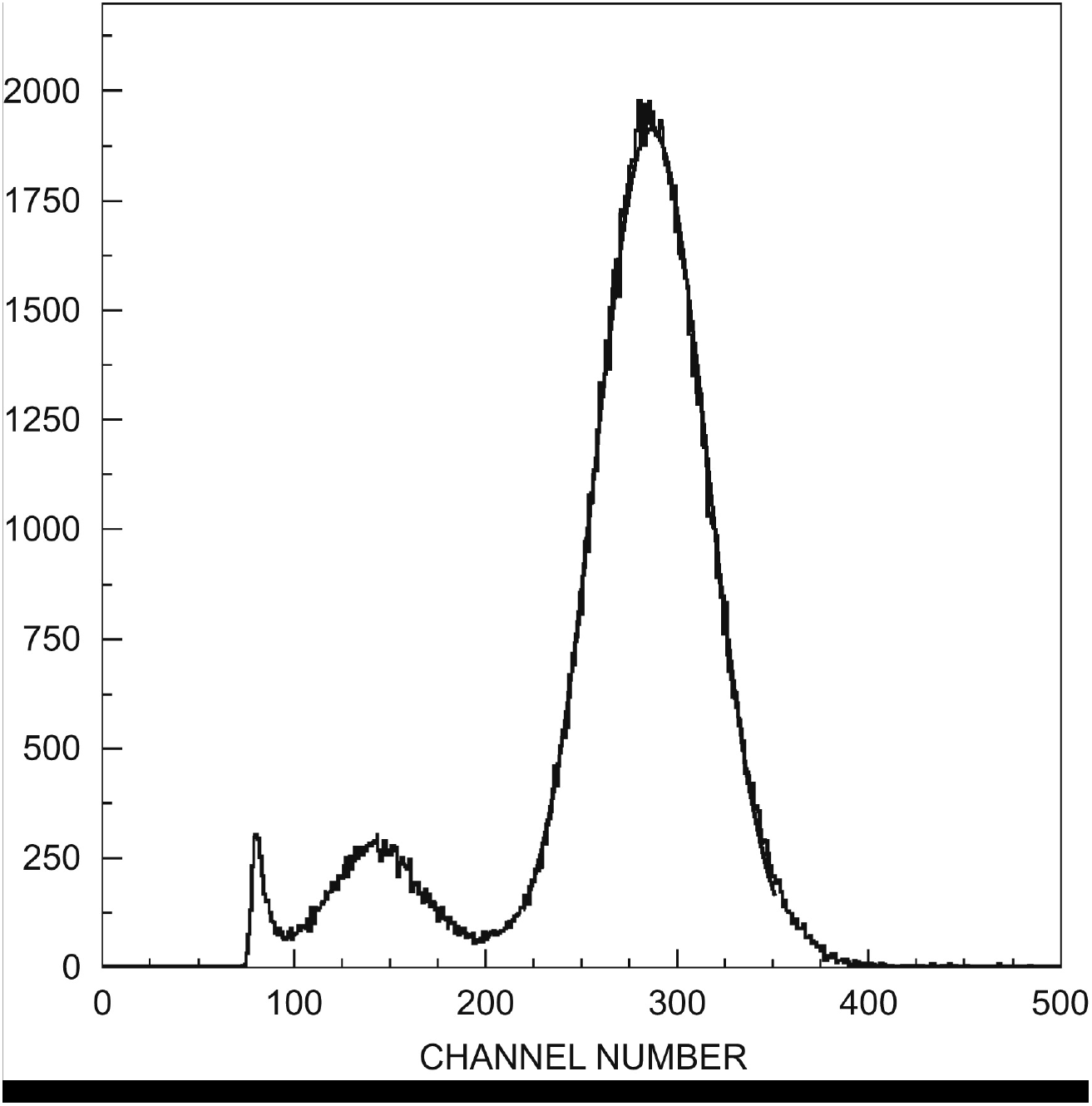}}
  	\caption{GEM simulation and $^{55}Fe$ spectrum used for gain measurement and resolution.}
  	\label{fig:gem}
\end{figure}

The GEMs, purchased from CERN, are 5 cm x 5 cm x 60 $\mu m$, and composed of a kapton layer between two thin sheets of copper.  This structure is then chemically pierced with a hole spacing of 140 $\mu m$ and $\sim$100\% active area~\cite{Bressan:1998ji}.  They reliably provide a gain of up to 300 in $Ar/CO_2$, however when operating the top GEM at 500 V and bottom GEM at 400 V we obtain a gain of $\sim$40000.  Figure~\ref{fig:gemsim} shows a simulated avalanche of a single electron with a GEM bias of 500 V.

The gain of the GEMs is measured by placing an $^{55}Fe$ source in the vessel.  The copper collection pad is used to collect the avalanche gain of the primary ionization from the source X-rays.  This signal is recorded with a pulse height analyzer that was calibrated using a pulse generator.  An example spectrum can be seen in Figure~\ref{fig:espec}.  The primary peak corresponding to the 5.9 keV X-ray emission has a full-width-at-half-max (FWHM) of approximately 24\%.  The escape peak is also well resolved.

\subsubsection{Cosmic Ray Data}
\begin{figure}[h]
 	\centering
  	\subfloat[Reconstructed cosmic track~\cite{Kim:2008zzi}.]{\label{fig:costrk}\includegraphics[trim = 1mm 0mm 0mm 0mm, clip,width=0.4\textwidth]{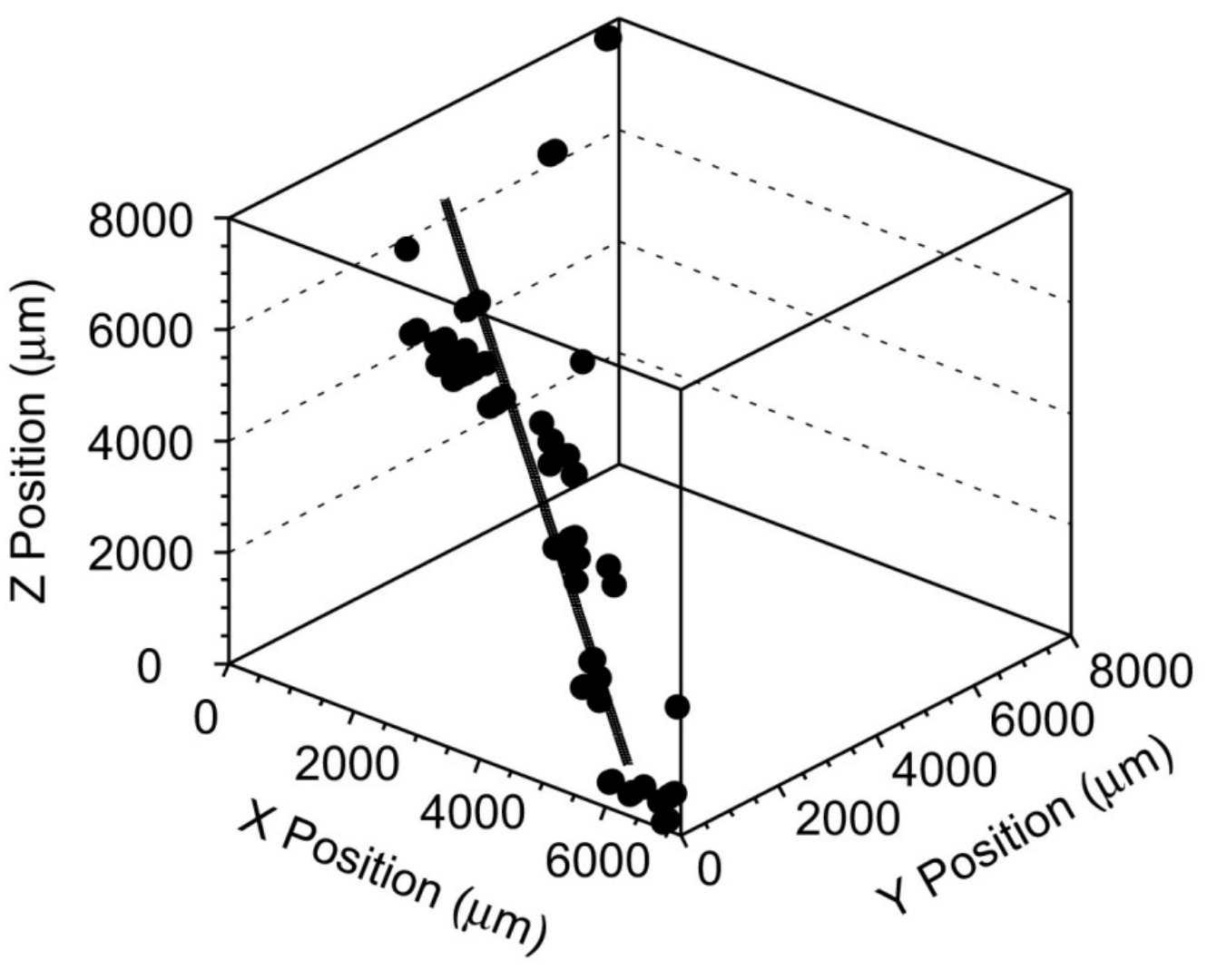}}                
	\hspace{0.1\textwidth}
  	\subfloat[Rate of cosmic ray detection as a function of GEM gain~\cite{Kim:2008zzi}.]{\label{fig:coseff}\includegraphics[width=0.4\textwidth]{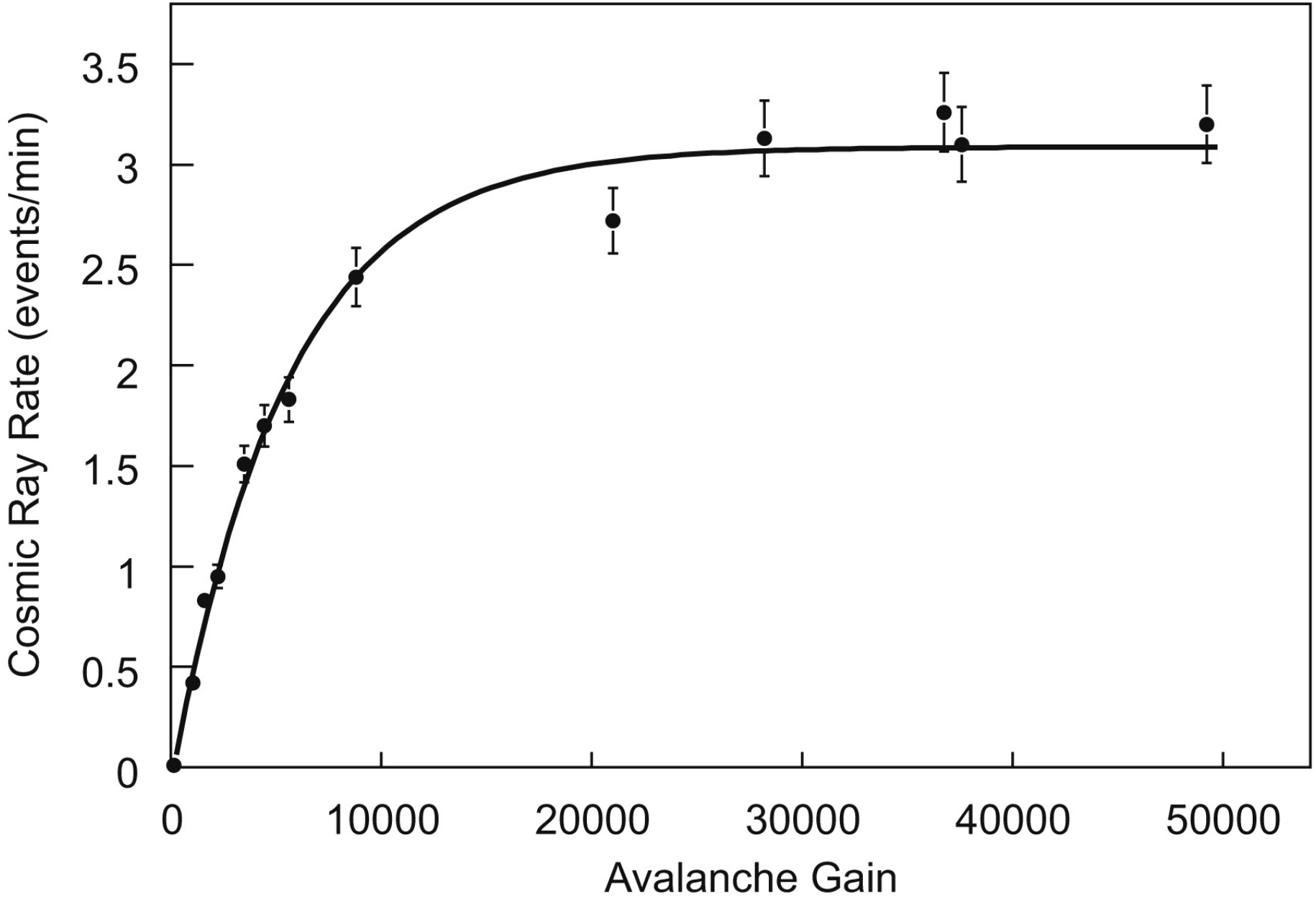}}
  	\caption{Cosmic ray data collected with the LBNL First Generation prototype.}
  	\label{fig:data}
\end{figure}

A large sample of cosmic ray events were collected with the LBNL prototype.   Figure~\ref{fig:costrk} shows one of these cosmic ray candidate track. For these events the drift field was set to 1 kV/cm, the gain was set to 9000, and the pixel threshold was set to 1800 electrons.  Table~\ref{tab:fit} shows the x-z-y results for the cosmic ray sample.  

We then changed the pixel threshold and varied the GEM gain to test the detection efficiency with cosmic rays.  Figure~\ref{fig:coseff} shows the detection rate as a function of gain for a pixel threshold of 5000 electrons.  The plot plateaus at a gain value consistent with a few times trigger threshold.  This is consistent with all electrons being detected from the cosmic tracks.

\begin{table}[h]
	\centering
	\begin{tabular}{| c | c  | c  | c |}
		\hline
	    		& Fit Residual ($\mu m$) & Diffusion ($\mu m$)& $ \sigma_{GEM+Pix} $ ($\mu m$) \\
		\hline \hline
		$ {x} $ & 170 & 110 & 130 \\  \hline
		$ {y} $ & 130 & 110 & 70 \\  \hline
		$ {z} $ & 240 & 190 & 150	 \\  \hline
	\end{tabular}		
	\caption{Fit results for cosmic ray tracks. The total GEM gain was 9000 with a minimum 10 pixel hits with pixel threshold set to 1800 electrons.  Each track was required to be greater than 4.5 mm.}
  	\label{tab:fit}
\end{table}	

\subsubsection{University of Hawaii, $D^3$ Micro}
\begin{figure}[h]
 	\centering
  	\subfloat[GEM and pixel fixture constructed at UH.]{\label{fig:uhfix}\includegraphics[width=0.4\textwidth]{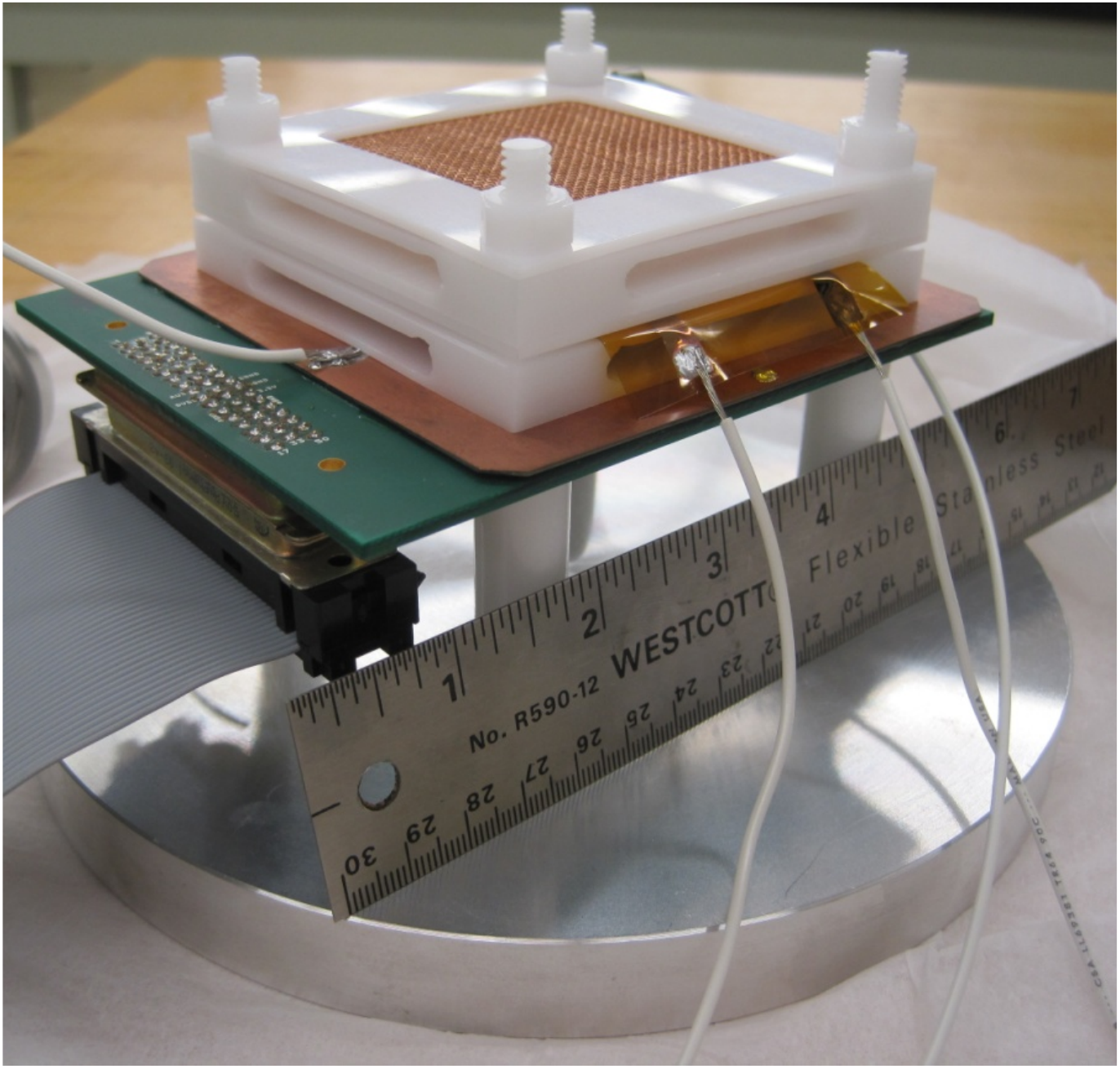}}                
	\hspace{0.05\textwidth}
  	\subfloat[Vacuum vessel at UH]{\label{fig:uhves}\includegraphics[width=0.45\textwidth]{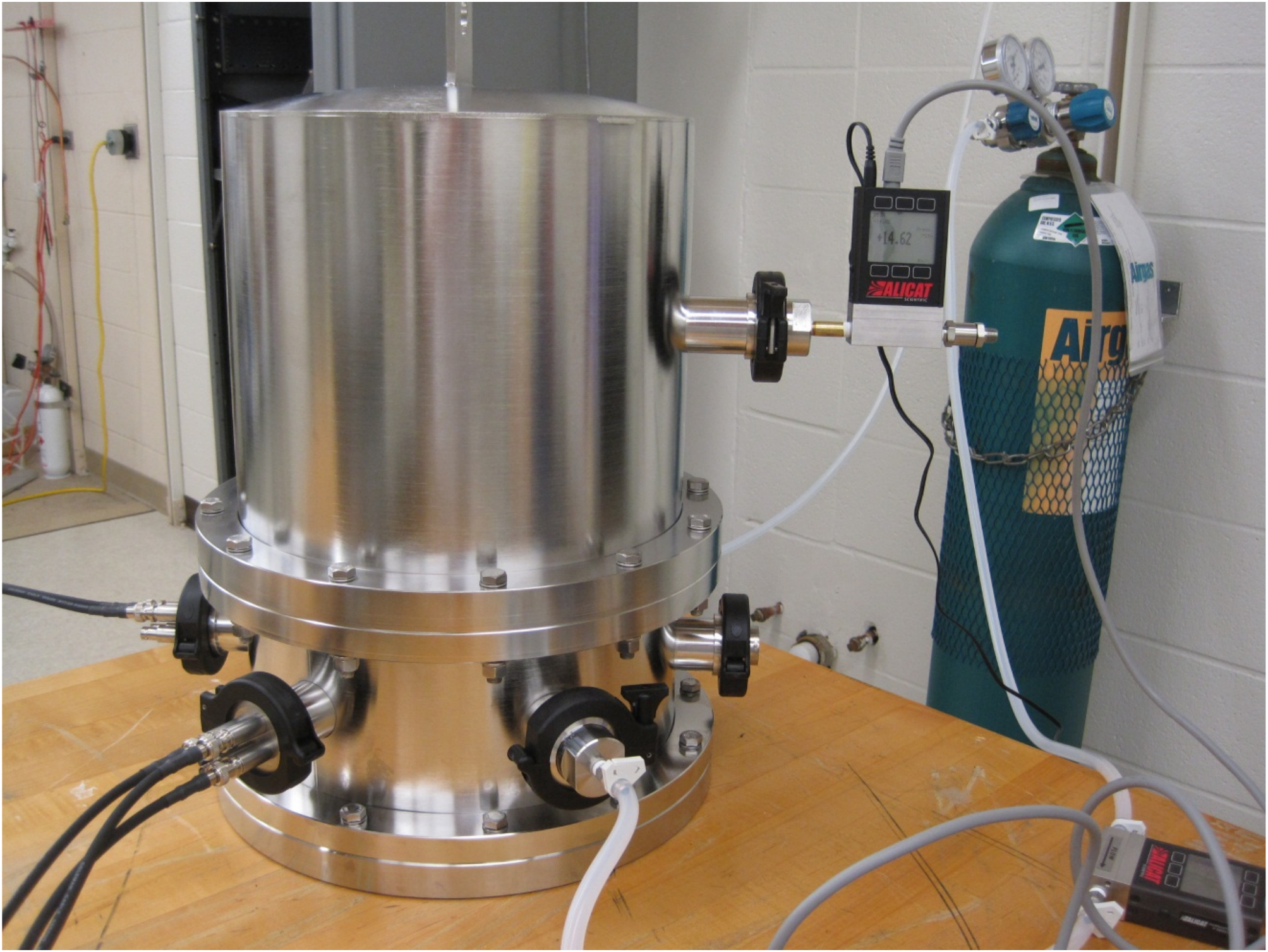}}
  	\caption{$D^3$ Micro detector being constructed at UH.}
  	\label{fig:uh}
\end{figure}
The University of Hawaii (UH) is currently constructing a duplicate of the LBNL prototype.  Figure~\ref{fig:uhfix} shows our GEM/pixel fixture, and Figure~\ref{fig:uhves} is our vacuum vessel.  We are near completion and have taken our first data using a $^{57}Co$ source.   

\subsection{Second Generation}
\begin{figure}[H]
 	\centering
  	\subfloat[Cross section of Second Generation prototype.]{\label{fig:schim2nd}\includegraphics[trim = 0mm 8mm 0mm 0mm, clip, width=0.4\textwidth]{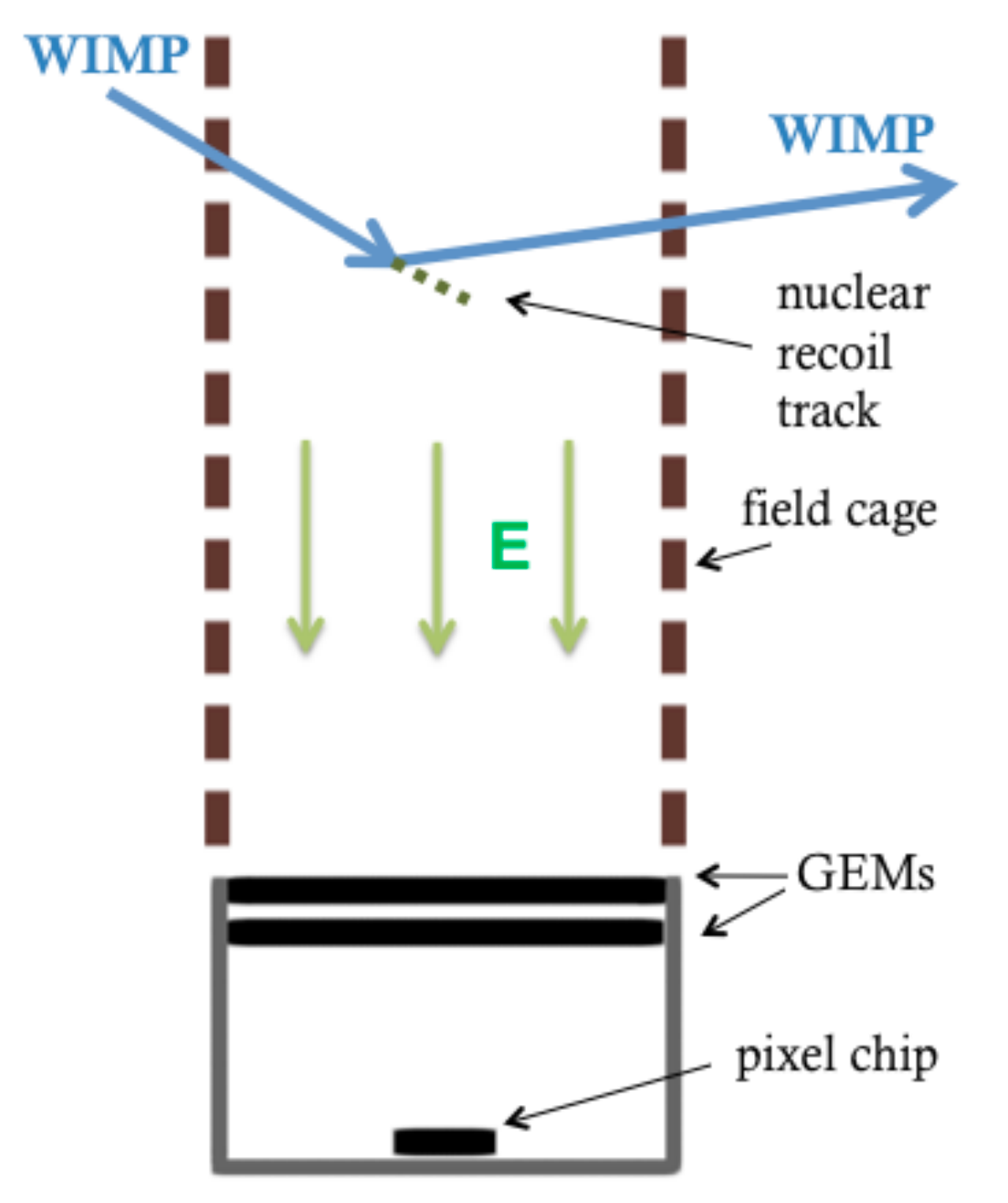}}                
	\hspace{0.1\textwidth}
  	\subfloat[Computer model of the Second Generation prototype, including field cage and GEM/pixel fixture.]{\label{fig:cad2nd}\includegraphics[width=0.2\textwidth]{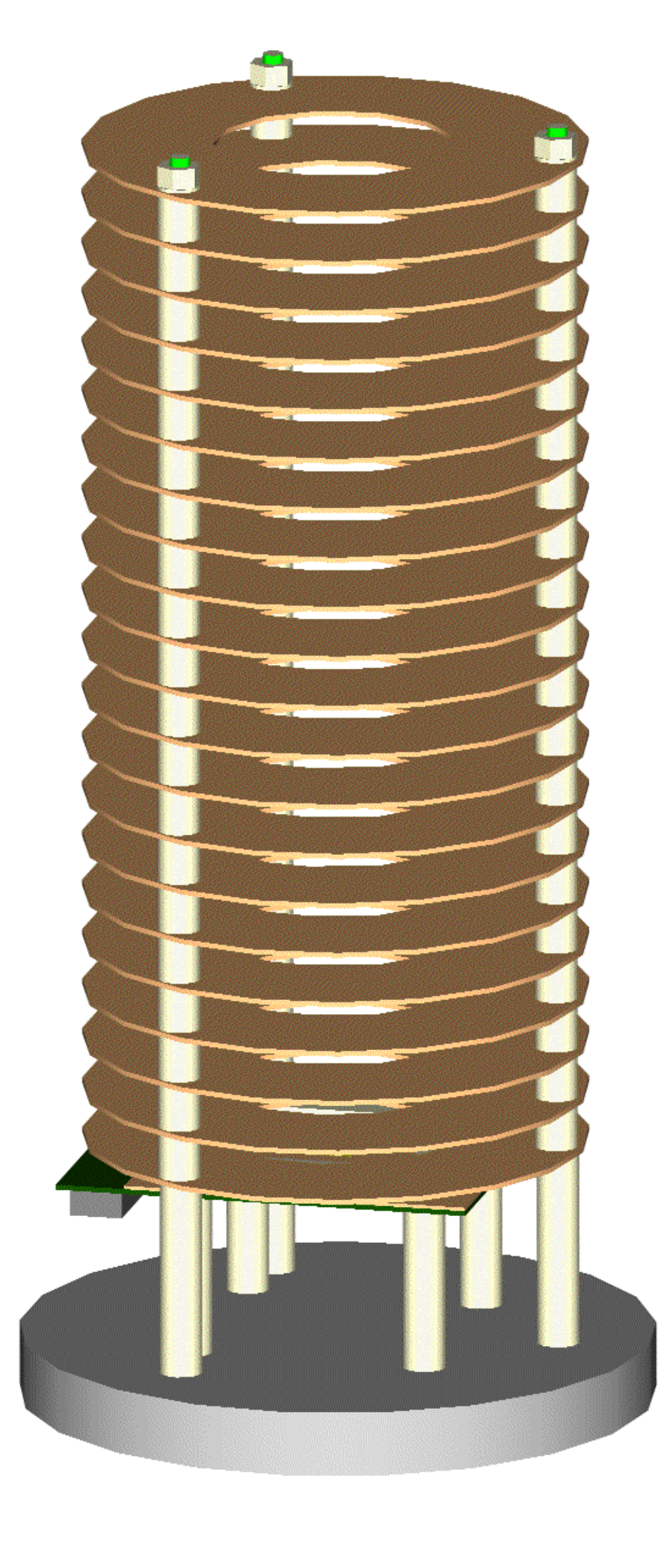}}
  	\caption{Proposed Second Generation prototypes.}
  	\label{fig:2ndGen}
\end{figure}

The Second Generation of the detector is a simple extension of the First Generation prototype.  The main limitation of the first prototype is its small active volume, thus we simply remove the drift electrode and replace it with a 30 cm field cage, Figure~\ref{fig:schim2nd}.  Figure~\ref{fig:cad2nd}, shows a computer aided design (CAD)~\cite{cadtext} drawing of the proposed Second Generation prototype.  During this phase of development we will begin to optimize the detector differently for dark matter searches ($D^3$) and for neutron detection (Directional Neutron Observer, DiNO).

\section{Future Plans}
\label{Fut_Plans}
\begin{figure}[H]
 	\centering
  	\subfloat[Four cell $D^3$ Milli/DiNO detector.]{\label{fig:dino}\includegraphics[width=0.45\textwidth]{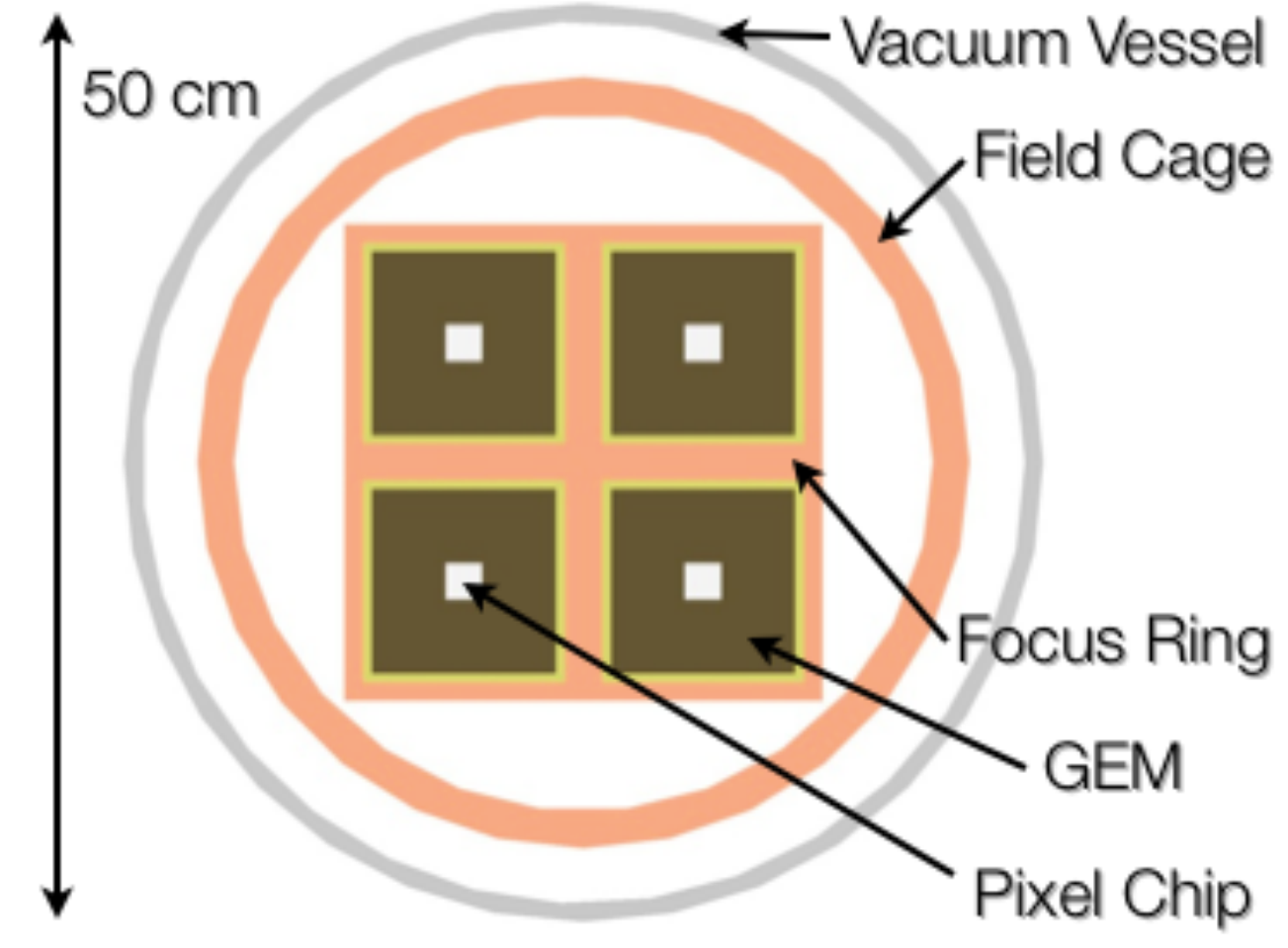}}                
	\hspace{0.1\textwidth}
  	\subfloat[Proposed $D^{3}$ detector, approximately 1.5 m in diameter]{\label{fig:d3}\includegraphics[width=0.35\textwidth]{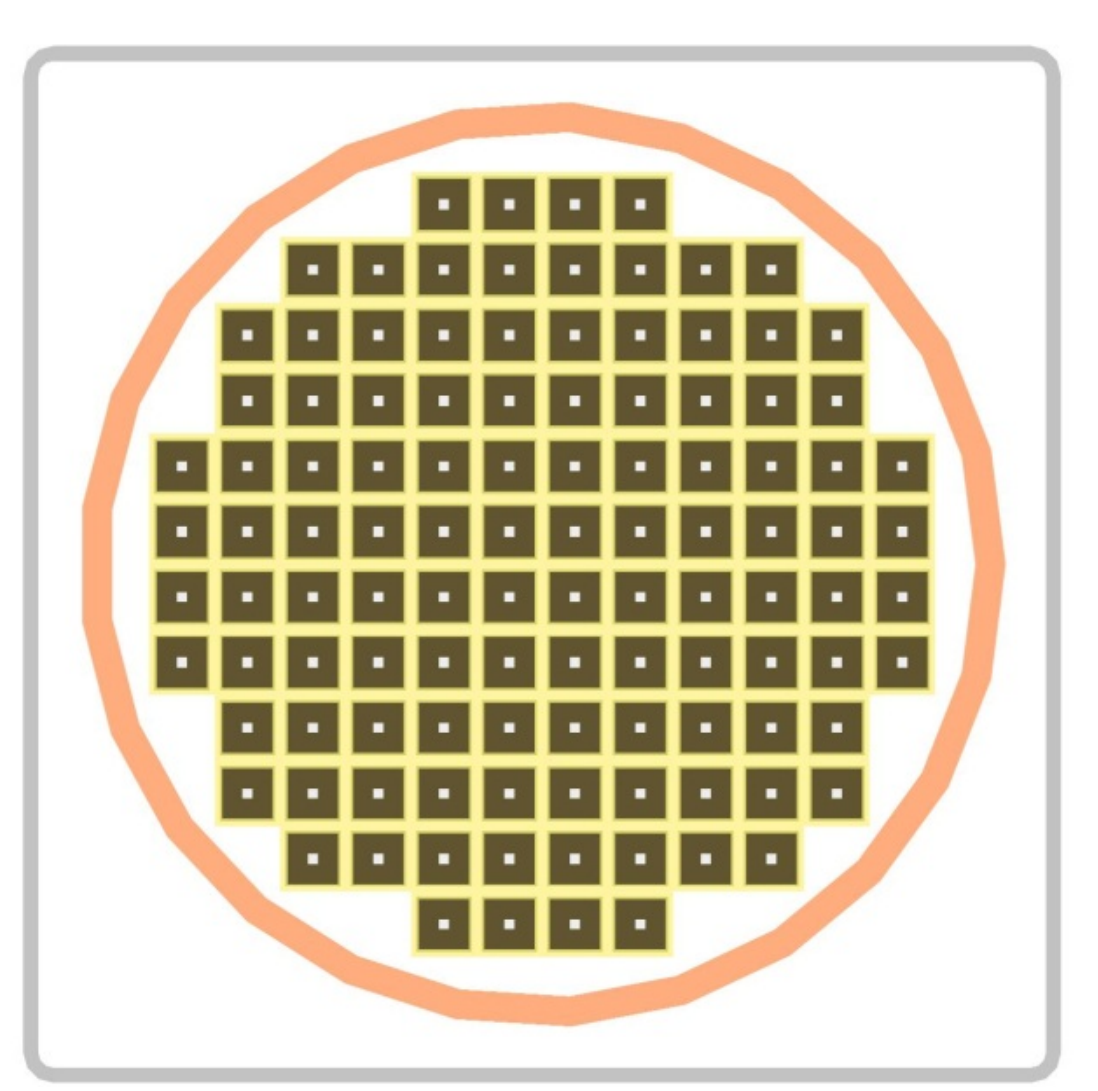}}
  	\caption{Proposed $D^3$ Milli and $D^3$ detectors.}
  	\label{fig:fut}
\end{figure}

The ATLAS collaboration has continued the development of pixel chips for future upgrades.  The next chip, FE-I4~\cite{GarciaSciveres:2011zz}, will be 2 cm x 2 cm.  It will be paired with larger GEMs, 10 cm x 10 cm, enabling instrumentation of a larger volume.  This will be implemented with the USBPIX readout developed at  Bonn University~\cite{website:usbpix}.  This USB based readout system will further simplify the readout of multiple pixel chips.  We are also working on a mechanism to focus the drifting electrons which will enable us to  instrument a large volume at a lower cost.   This work will be published elsewhere but is currently incorporated in the design of $D^3$/DiNO.  

With all (or even some) of these improvements, and a larger detector we may be able to detect dark matter.  Figure~\ref{fig:fut} shows the proposed cellular design to be used in $D^3$/DiNO, where each cell is similar to Figure~\ref{fig:schim2nd}.  Several layers, constructed as in Figure~\ref{fig:d3}, can be used to increase the active volume. 
%
%
\begin{figure}[H]
\centering
\includegraphics[width=0.7\textwidth]{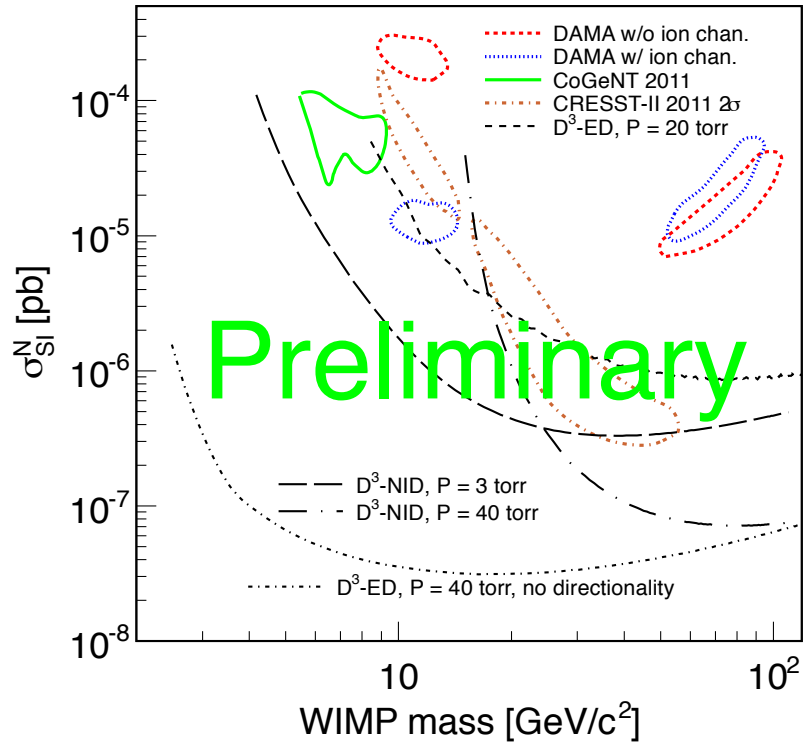}
\caption{Limit plot for $D^3$.  Limits set using negative ion drift (NID) and electron drift (ED).  Also shown are results for CRESST-II~\cite{Angloher:2011uu},  DAMA/LIBRA~\cite{Bernabei:2010mq}, and CoGeNT~\cite{Aalseth:2011wp}}
\label{fig:limit}
\end{figure}

Figure~\ref{fig:limit}, shows the expected limit for several different running scenarios.  This plot includes a scenario where we use ion drift rather than electron drift~\cite{2005NIMPA.555}.  This seems quite promising because of the reduction in diffusion, but we have not explored this experimentally yet.  The electron drift (ED) limits and negative ion drift (NID) limits are calculated for $CF_4$ and $CS_2$ respectively.  Both assume a one year of exposure to a detector with three $m^3$ of active volume constructed of nine layers, each with a 33.33 cm drift distance.  This shows we may have sensitivity to areas of interest found by  CRESST-II~\cite{Angloher:2011uu},  DAMA/LIBRA~\cite{Bernabei:2010mq}, and CoGeNT~\cite{Aalseth:2011wp}.  While our ability to run at these low pressures have yet to be demonstrated, this study does suggest our possible discovery potential.  Further information on our simulations can be found in~\cite{igal:cygnus}.  




\section{Conclusions}
\label{Conclusions}
While TPCs have been used for decades in high energy physics, we are developing a new variation that utilizes their strengths, large detector volume, and increases the precision and efficiency with the use of pixels and GEMs.  The LBNL prototype has shown the detector design works well with cosmic rays.  We have outlined a clear path to a larger detector, and have shown via simulation we may have sensitivity to discover or exclude a wide range of dark matter models. 

Other projects, like the DMTPC collaboration~\cite{Sciolla:2008ak}, are using TPC for directional dark matter searches.  We believe our use of the FE-I3 pixel chip will provide superior spacial and energy resolution.  Since the TIPP conference, the University of Hawaii group has joined the DMTPC collaboration with hope of leveraging the expertise of both groups.






\bibliographystyle{elsarticle-num}
\bibliography{dcube}







\end{document}